\documentclass[showpacs,aps,prd,nofootinbib,tightenlines,nobibnotes,floatfix,amsmath,amssymb,11pt]{revtex4-1}
\usepackage{graphicx}
\usepackage{color}
\usepackage{enumerate} 
\usepackage{multirow}

\begin{document}

\title{Probing the  $h c\bar{c} $ coupling at a Future Circular Collider \\[0.15cm]
in the electron-hadron mode }

\author{J. Hern\'andez-S\'anchez}
\email{jaime.hernandez@correo.buap.mx}
\affiliation{Fac. de Cs. de la Electr\'onica, Benem\'erita Universidad Aut\'onoma de Puebla, Apartado
 Postal 1152, 72570 Puebla, Puebla, M\'exico and Dual C-P Institute of High Energy Physics, M\'exico}
\author{C. G. Honorato }
\email{carlosg.honorato@correo.buap.mx}
\affiliation{Fac. de Cs. de la Electr\'onica, Benem\'erita Universidad Aut\'onoma de Puebla, Apartado
 Postal 1152, 72570 Puebla, Puebla, M\'exico}
\author{S. Moretti}
\email{s.moretti@soton.ac.uk}
\affiliation{School of Physics and Astronomy, University of Southampton, Highfield, Southampton SO17
 1BJ, United Kingdom and Particle Physics Department, Rutherford Appleton Laboratory, Chilton, Didcot,
 Oxon OX11 0QX, United Kingdom}


\date{\today}

\begin{abstract}
We study the production of a neutral Higgs boson at a Future Circular Collider in the electron-hadron mode (FCC-eh) through the leading process 
$e^- p \to \nu_e h q  $ assuming the decay channel $h \to c \bar{c} $, where $h$ is  the Standard Model (SM)-like state discovered at the Large Hadron Collider (LHC). This process is studied in the context of  a 2-Higgs Doublet Model Type III (2HDM-III) embedding a four-zero texture in the Yukawa matrices and a general Higgs potential, where  both Higgs doublets are coupled with up- and down-type fermions.  Flavour Changing Neutral Currents (FCNCs) are well controlled  by this approach through the adoption of a suitable texture once flavour physics constraints are taken in account.     
Considering the parameter space where the signal is enhanced and in agreement with both experimental data and theoretical conditions, we analyse the aforementioned signal by taking into account the most important SM backgrounds, separating $c$-jets from light-flavour  and gluon ones as well as $b$-jets by means of efficient flavour tagging. We find that the $h c \bar{c} $  coupling strength can be accessed with good significance after a  luminosity of 
$1$ ab$^{-1}$ for  a 50 TeV proton beam and a 60 GeV electron one, the latter with a 80\% (longitudinal) polarisation.
  
\end{abstract}

\maketitle

\section{Introduction}

Since the July 2012 discovery of a Higgs boson, $h$, with properties very consistent with those predicted by the Standard Model (SM), at the Large Hadron Collider (LHC), by the 
ATLAS and CMS experiments, spontaneous Electro-Weak Symmetry Breaking (EWSB) resting on a minimal Higgs mechanism  is apparently well established. While a mass of 125 GeV is not really an indication for the SM construct being the one responsible for these LHC signals, as  $m_h$ is a free parameter in it, the fact that production and decay rates involving couplings to $W^\pm$ and $Z$ bosons as well as $t,b,\tau$ and $\mu$ fermions have been measured and are compatible with the SM genuine predictions (once $m_h$ is measured) is a strong sign in flavour of such a minimal Higgs construct.

However, a notable absence in the list of the SM-like Higgs couplings so far measured is the one involving the $hc\bar c$ vertex, which is presently unconstrained. The reason is that corresponding signals at the LHC are masked by an enormous QCD background. In fact, not even the ability to tag $c$-jets, on a similar footing with what has successfully been done for $b$-jets, is sufficient to enable a measurement of the Yukawa coupling to $c$-(anti)quarks at a level comparable to the case of $b$-ones, as the displaced vertex associated to semi-leptonic $c$-meson decays is much closer to the interaction point (where the gluon and light flavour jet backgrounds originates)     than the one stemming from the corresponding $b$-meson transitions. Another drawback is that the strength of the $hc\bar c$ vertex in the SM is much smaller than that of the $hb\bar b$ vertex, as they scale with the fermion mass, which in turn means that the Branching Ratios (BRs) for $h\to c\bar c$ is $(m_c/m_b)^2$ times smaller than BR($h\to b\bar b$), see, e.g., Ref.~\cite{Moretti:1994ds}.  On the other hand,  the precise evaluation of  decay width  $h \to c \bar c$ has been studied recently at next-to-next-to-leading order QCD (including the flavour-singlet contribution) and the next-to-leading order electroweak \cite{Li:2021ygc}.

However, there exist extensions of the SM in which the Higgs sector is enlarged by additional (pseudo)scalar multiplets (singlets, doublets, triplets, etc.), where the $h\to c\bar c$ rate can be increased substantially. Herein, owing to the fact that the SM-like Higgs discovered at the LHC has a clear doublet nature, we focus  on the simplest extension of the SM involving such Higgs multiplet, the so-called  
2-Higgs Doublet Model (2HDM). The latter comes in several guises, known as  Type I, II, III (or Y)   and IV (or X), wherein Flavour Changing Neutral Currents (FCNCs) mediated by (pseudo)scalars can be eliminated under discrete symmetries \cite{Branco:2011iw,Barger:1989fj,Grossman:1994jb,Aoki:2009ha,Gunion:1989we,Donoghue:1978cj,Barnett:1984zy}, entirely if the latter are exact or sufficiently to comply with experimental limits if they are softly broken.  In fact, another, very interesting kind of  2HDM is  the one where FCNCs can be controlled by a particular texture in the Yukawa matrices \cite{Fritzsch:2002ga}. In particular, in previous papers, we have implemented a four-zero texture, in a scenario which we have called 2HDM Type III (2HDM-III) \cite{DiazCruz:2009ek}. This model has a phenomenology that is very rich, which we studied at colliders in various instances \cite{Hernandez-Sanchez:2016vys,HernandezSanchez:2013xj},
and some very interesting aspects, like flavour-violating quarks decays, which can be enhanced for  neutral Higgs bosons with intermediate mass (i.e., below  the top quark mass). {In particular, we have studied the signal $\phi_i^0 \to s \bar{b} + h.c.$ ($\phi_i^0= h, \, H$) \cite{Hernandez-Sanchez:2015bda, Das:2015kea}.  Furthermore, in this model, the parameter space can avoid the current experimental constraints from flavour and Higgs physics and   a light charged Higgs  boson is allowed \cite{HernandezSanchez:2012eg}, so that  the  decay $H^- \to b \bar{c}$ is enhanced and its 
BR can be dominant \cite{Akeroyd:2016ymd,Akeroyd:2012yg}.  

In fact, the 2HDM-III is also an ideal candidate in providing enhanced $h\to c\bar c$ rates, as the Yukawa texture parameters that affect the aforementioned $H^\pm$ signatures also enter the 
 $h\to b\bar b$ and $h\to c\bar c$ ones. In particular, it is always possible to maintain the Yukawa coupling to the $b$-(anti)quark  in the range currently measured at the LHC, and indeed the one foreseen by the end of the High Luminosity LHC (HL-LHC) era \cite{Gianotti:2002xx,Apollinari:2017lan}, while enhancing the $h\to c\bar c$ one. It is the purpose of this paper to study the scope of the  electron-proton Future Circular Collider (FCC-eh), with a Center-of-Mass (CM) energy of 3.5 TeV \cite{Kuze:2018dqd,Britzger:2017fuc,Abada:2019lih}. 
%%%%%%%%%%%%%%%%%%%%
 This configuration is 
obtained by the collisions of  a 50 TeV proton beam coming from the FCC-hh  \cite{FCC:2018vvp}  and a 60 GeV electron beam from an external linear accelerator (Electron Recovery Linac (ERL)) tangential to the FCC main tunnel  \cite{Abada:2019lih}   
%%%%%%%%%%%%%%%%%%%%%%%%
and offers good prospects as a Higgs boson factory, 
as herein one could elucidate the nature of the couplings of generic Higgs bosons  to most fermions, especially the $h \to b \bar{b} $ one, which remains difficult to establish with high precision at both the LHC and the HL-LHC because of the overwhelming QCD background \footnote{Other interesting studies for probes of Higgs coupling in  Higgs pair production at hadron electron colliders LHeC and FCC-eh have been realised recently \cite{Jueid:2021qfs}. }.  Given these encouraging results for the $hb\bar b$ vertex, we specifically analyse here   
the prospects of also establishing the $hc\bar c$ one.

Our work is organised as follows. In the next section we describe briefly the 2HDM-III, specifically, its Yukawa structure. Then in the following one we discuss the theoretical and experimental constraints applying to it and select some benchmark scenarios for numerical analysis. In  section \ref{sec:results} we give our  results whereas in section \ref{sec:conclusions} we finally summarise.

\section{The Higgs sector of the 2HDM-III}

The 2HDM-III is described by two scalar Higgs doublets  $\Phi_i =(\phi_i^- , \, \phi_i^{0*} ) $ ($i=1$, 2), with hypercharge $+1$, which can  couple to all fermions. FCNCs are controlled by a specific four-zero texture in the Yukawa matrices, the latter being an effective flavour theory of the Yukawa sector. Therefore, a discrete symmetry is not necessary in this approach, so that the $SU(2)_L \times U(1)_Y$  invariant scalar potential in its general form can be considered, which is given by \cite{ Haber:1993an,Gunion:2002zf,Dubinin:2002nx,Gunion:2005ja}:
%%%%%%%%%%%%%%%%%%%%%%%%%%%
\begin{eqnarray}
V(\Phi_1,\Phi_2)&=&\mu_1^2(\Phi_1^\dagger \Phi_1)+\mu_2^2(\Phi_2^\dagger \Phi_2)-\Big(\mu_{12}^2(\Phi_1^\dagger \Phi_2+h.c.)\Big)\nonumber\\
&&+\frac{1}{2}\lambda_1(\Phi_1^\dagger \Phi_1)^2+\frac{1}{2}\lambda_2(\Phi_2^\dagger \Phi_2)^2+\lambda_3(\Phi_1^\dagger\Phi_1)(\Phi_2^\dagger\Phi_2)+\lambda_4(\Phi_1^\dagger\Phi_2)(\Phi_2^\dagger\Phi_1)\nonumber\\
&&+\left(\frac{1}{2}\lambda_5(\Phi_1^\dagger\Phi_2)^2+\lambda_6(\Phi_1^\dagger\Phi_1)(\Phi_1^\dagger\Phi_2)+\lambda_7(\Phi_2^\dagger\Phi_2)(\Phi_1^\dagger\Phi_2)+h.c.\right),
\end{eqnarray}
%%%%%%%%%%%%%%%%%%%%%%%%%%%%%%%
wherein, for simplicity, we  suppose that all parameters are real\footnote{The $ \mu_{12}^2 $ and $ \lambda_i$ $ (i=5,6,7)$ parameters could be complex in general, which then induce  CP-violation in the Higgs sector.} as so are the Vacuum Expectation Values (VEVs) of the Higgs fields.   	Besides, notice that, when a discrete symmetry is implemented in the model, the $\lambda_{6,7}$ terms are absent. However, in our model, the latter can be kept in the Higgs potential when the four-zero texture is implemented in the Yukawa matrices.   This is rather interesting, as we have shown that these parameters ($\lambda_{6,7}$) can be relevant in one-loop processes  but do not contribute to EW parameter   $\rho = m_W^2/m_Z^2 \cos_W^2$ \cite{Cordero-Cid:2013sxa}.  However, the ordinary  custodial symmetry \cite{Branco:2011iw,Gunion:1989we}(twisted custodial symmetry \cite{Gerard:2007kn})  associated to the $\rho$ parameter is broken when the difference $m_{H^\pm} -m_A $($m_{H^\pm} -m_H$) is sizeable, being $H^\pm$ the charged Higgs boson and $A(H)$ the heavy CP-odd(even) one
belonging to this construct, in addition to the aforementioned SM-like $h$ state. Reasonable models with such an extended Higgs sector are those for which $\rho \approx 1$ when radiative corrections are included \cite{Toussaint:1978zm,Bertolini:1985ia,Hollik:1986gg,Gunion:1988pc,Gunion:1989we,Gunion:2002zf,Gerard:2007kn,deVisscher:2009zb} or, more in general, those in good agreement with the experimental constraints from the oblique parameters $S$, $T$ and $U$ \cite{Haller:2018nnx,Kanemura:2011sj}, part of the so-called EW Precision Observables (EWPOs) \cite{pdg:2020}. The described Higgs bosons spectrum emerges after EWSB, which provide mass to the $W^\pm$ and $Z$ bosons, thus releasing five physical Higgs fields:  two CP-even neutral states  $h,H$ (with $m_h < m_H$),  one CP-odd neutral state $A$ plus two charged Higgs bosons $H^\pm$. Furthermore, one also has the mixing angle $\alpha$, that relates the two CP-even neutral bosons $(h,H) $ and $\beta$ (being $\tan \beta$ the ratio of VEVs of the two Higgs doublets). The masses of these Higgs fields  and these two angles are the inputs parameters chosen here to describe the scalar potential. 

About the Yukawa sector of our model, this is defined by  \cite{HernandezSanchez:2012eg}:
%%%%%%%%%%%%%%%%%%%%%%%%%%%%%%%%
\begin{equation}
\label{yuklan}
{\cal{L}}_Y=-\left(Y_1^u\bar{Q}_L\tilde{\Phi}_1u_R+Y_2^u\bar{Q}_L\tilde{\Phi}_2u_R+Y_1^d\bar{Q}_L\Phi_1d_R
                                +Y_2^d\bar{Q}_L\Phi_2d_R+Y_1^l\bar{L}_L\tilde{\Phi}_1l_R+Y_2^l\bar{L}_L\tilde{\Phi}_2l_R\right),
\end{equation}
%%%%%%%%%%%%%%%%%%%%%%%%%%
being $\tilde{\Phi}_i = i \sigma_2 \Phi_{i}^*$ ($i=1, 2$) and where both Higgs doublets are coupled with up- and down-type fermions. Following the procedure of Refs. \cite{Felix-Beltran:2013tra, HernandezSanchez:2012eg}, after  EWSB, the fermion mass matrices are:
\begin{eqnarray}
M_f=\frac{1}{\sqrt{2}}(v_1Y_1^f+v_2Y_2^f)\ \ \ (f=u,d,\ell),
\end{eqnarray}
where the Yukawa matrices $Y_{1,2}^f$ have the four-zero texture form and are Hermitian. Considering the diagonalisation of the fermion mass matrices through $\bar{M}_f= V_{fL}^\dag M_f V_{fL} $, we have $\tilde{Y}_n^f= V_{fL}^\dag Y_n^f V_{fL}$, then one can get a good approximation for the rotated matrix  $\tilde{Y}_n^f$ as follows  \cite{HernandezSanchez:2012eg}:
\begin{eqnarray}
\Big(\tilde{Y}_n^f\Big)_{ij}=\frac{\sqrt{m_i^fm_j^f}}{v}(\tilde{\chi}^f)_{ij}= \frac{\sqrt{m_i^f m_j^f}}{v}\chi_{ij}^f e^{\theta_{ij}^f},
\end {eqnarray}
where  the  $\chi$s are dimensionless and constrained by  flavour physics experimental data, which will be discussed in the following section. Then, one can obtain the generic Lagrangian of the Yukawa sector, which gives the  interactions of physical (pseudo)scalars fields with fermions,   as\footnote{One can assume this Lagrangian is the one of an effective field theory, wherein the Higgs fields play a relevant role in the flavour structure of some high scale renormalisable flavour model \cite{Aranda_2012, Branco_2010, Botella_2013, Frigerio:2004jg}.}:
%%%%%%%%%%%%%%%%%%%%%%%%%%%%%%%%%
\begin{eqnarray}
{\cal{L}}^{\bar{f}_if_j\phi}&=&-\left\{\frac{\sqrt{2}}{v}\bar{u}_i\big(m_{d_j}X_{ij}P_R+m_{u_i}Y_{ij}P_L\big)d_jH^+ +\frac{\sqrt{2} m_{l_j}}{v}Z_{ij}\bar{\nu}_Ll_RH^++h.c.\right\}\nonumber\\
&&-\frac{1}{v}\left\{\bar{f}_i m_{f_j}h_{ij}^ff_jh^0-i\bar{f}_im_{f_i}A_{ij}^ff_j\gamma_5A^0\right\},
\end{eqnarray}
where $X_{ij}$, $Y_{ij}$, $Z_{ij}$, $A_{ij}$ are given in  \cite{HernandezSanchez:2012eg}. For our study it is sufficient to consider the functions $h_{ij}^f$, which are given by:
\begin{eqnarray}
h_{ij}^u&=&\varepsilon_h^u\delta_{ij}-\frac{\big(\varepsilon_H^u+Y\varepsilon_h^u\big)}{\sqrt{2}f(Y)}\sqrt{\frac{m_{u_j}}{m_{u_i}}}\chi_{ij}^u,\\
h_{ij}^d&=&\varepsilon_h^d\delta_{ij}+\frac{\big(\varepsilon_H^d-X\varepsilon_h^d\big)}{\sqrt{2}f(X)}\sqrt{\frac{m_{d_j}}{m_{d_i}}}\chi_{ij}^d,
\end{eqnarray}
wherein the parameters $X$, $Y$, $\varepsilon_{h,H}^{u,d}$ are given in  Table \ref{couplings}, where it is made clear that our 2HDM-III construct with a four-zero Yukawa texture can be related to the ordinary Yukawa types known as Type I, II and Y (or flipped) by choosing appropriately these parameters.

\begin{table}[htp]
\begin{center}
\begin{tabular}{|c|c|c|c|c|}
\hline
\hline
 2HDM-III &  $X$  &  $Y$  & $\varepsilon_h^u$ &  $\varepsilon_h^d$   \\
 \hline
 2HDM-I&  $-\cot\beta$ & $ \cot\beta$ & $\cos \alpha / \sin \beta$ &   $ \cos \alpha / \sin \beta$ \\
 \hline 
  2HDM-II&  $\tan\beta$ & $ \cot\beta$ & $\cos \alpha / \sin \beta$ &   $ -\sin \alpha / \cos \beta$ \\
  \hline 
  2HDM-Y  (flipped)&  $\tan\beta$ & $ \cot\beta$ & $\cos \alpha / \sin \beta$ &   $ -\sin \alpha / \cos \beta$ \\
 \hline
\hline
\end{tabular}
\end{center}
\caption{The parameters choices for $X$, $Y$ and $\varepsilon_h^{u,d}$ of the 2HDM-III needed to obtain the standard 2HDM Type I, II and Y (flipped),  which happen when the $\chi$s are zero.  }
\label{couplings}
\end{table}%
In general, the Higgs-fermion-fermion couplings are expressed as $g^{ff\phi}_{\rm 2HDM-III} = g^{ff\phi}_{\rm any}+ \Delta g$, where $g^{ff\phi}_{\rm any}$ represents the  $ff \phi$ coupling in any 2HDM with a discrete symmetry and $\Delta g$ is the  contribution of the four-zero texture  
\cite{ HernandezSanchez:2012eg}.

\section{Constraints and Benchmark Scenarios } 

In previous works \cite{DiazCruz:2009ek,Felix-Beltran:2013tra, HernandezSanchez:2012eg, Cordero-Cid:2015zma}, we have constrained our model by   considering EWPOs,  flavour and Higgs physics constraints from experimental data as well as theoretical bounds (such as unitarity\cite{Casalbuoni:1986hy, Ginzburg:2005dt}, vacuum stability \cite{Ferreira:2004yd,Ivanov:2007de} and perturbativity). Since the theoretical bounds and EWPO constraints have been analysed  very recently \cite{Hernandez-Sanchez:2020vax},  and these have not changed, we refer the reader to such a paper.  In contrast, experimental constraints evolve continuously so here we have re-evaluated them in the light of the very latest results. Specifically, 
the parameter space of the model was constrained by flavour physics measurements, through the  experimental data  bounds from leptonic and semileptonic meson decays, the inclusive decay   $B \to X_s \gamma$ , the $B_0-B_0$ and $ K_0 -K_0$ mixing as well as the process $B_s \to \mu^+ \mu^-$  \cite{HernandezSanchez:2012eg, PhysRevD.87.094031}.   We have then used HiggsBounds \cite{pdg:2020,ATLAS:2019nkf} and HiggsSignals \cite{ATLAS:2016neq, CMS:2018uag} to place  bounds over the masses and couplings of neutral \cite{CMS:2019ogx,CMS:2016xnc} and charged Higgs bosons \cite{ALEPH:2013htx,D0:1999rfq,CDF:2005acr,D0:2009oou,CMS:2018dzl,CMS:2019bfg},  so as to make sure that  the parameter space of the 2HDM-III considered here is consistent with any Higgs boson searches and measurements conducted at the LHC and previous colliders. In particular, 
using LHC measurements of the SM-like Higgs boson in the decays $h\to \gamma \gamma$ and $\gamma Z $  \cite{Aaboud:2018wps,Aaboud:2018ezd,Aad:2020plj,Aaboud:2017uhw,Sirunyan:2018koj,Sirunyan:2018tbk},  we made sure that the Yukawa texture  involving  the couplings of the charged Higgs boson with fermions in the  loops of these processes is in agreement with data. Specifically,  in the permitted region of parameter space of our model,  rather low masses  for the charged Higgs bosons are allowed \cite{Cordero-Cid:2013sxa,Cordero-Cid:2015zma,Hernandez-Sanchez:2020vax, Flores-Sanchez:2018dsr}.  

As intimated, in this study, the scalar field $h$  is the SM-like Higgs boson, hence,   $m_h =125$ GeV. Furthermore,  we choose the following parameter space: $m_A = 150$ GeV,  125 GeV $< m_H<$ 200 GeV, 100 GeV  $< m_{H^\pm} < 170$ GeV and $\cos (\beta -\alpha) \approx 0.1$, 
 with $0.014 \leq S \leq 0.026$ and $ -0.02 \leq T \leq 0.028$  and tensioned these against the measured values $S = -0.01 \pm 0.07$ and $T = 0.04 \pm 0.06$ (fixing $U = 0$) taken from \cite{Workman:2022ynf},
being all this parameter space  consistent with the aforementioned theoretical conditions  and experimental data.
Then, we select some sets of free parameters $\chi$'s which will represent our benchmark scenarios for each of the discussed 2HDM-III realisations  (or incarnations).  Explicitly, these benchmark scenarios are shown in Table  \ref{scenarios}.
\begin{table} 
\begin{center}
\begin{tabular}{|c|c|c|c|c|}
\hline 
Scenario & $\cos(\beta-\alpha)$ & $\chi^u_{\{22,23,33\}}$ & $\chi^d_{\{22,23,33\}}$ & $\chi^l_{\{22,23,33\}}$   \\ 
\hline
\hline 
Ia & 0.1 &$\{1,0.1,1.4\}$ &$\{1.8,0.1,1.2\}$ & $\{-0.4,0.1,1\}$ \\ 
\hline
IIa & 0.1 & $\{1,-0.53,1.4\}$ & $\{1.8,0.2,1.3\}$ & $\{-0-4,0.1,1\}$   \\ 
\hline
Y & 0.1 & $\{1,-0.53,1.4\}$ & $\{1.8,0.2,1.3\}$ & $\{-0-4,0.1,1\}$   \\ 
\hline 
\end{tabular}
\caption{Values for free parameters which define our benchmark scenarios, all being consistent with current theoretical and experimental bounds. }
\label{scenarios}
\end{center}
\end{table}
 We have characterised these benchmark scenarios in  Figure \ref{bench}, where  we show the events rates for
the aforementioned production and decay process
 over the $(X,Y)$ plane for case Ia, IIa and Y of Table \ref{scenarios}. These events rates are realised at parton level, taking the efficiency of $c$-tagged jets as $\epsilon_c=0.24$ and assuming 1 ab$^{-1}$ of (integrated) luminosity.  The coloured regions over the  $(X,Y)$ plane shown herein are compliant with  all aforementioned constraints while the white backgrounds correspond to 
regions ruled out. Furthermore,  we have demanded that for all benchmark points  the BR$(h\to b\bar b)$ is in agreement with the latest experimental observations, which established as ratio of the measured value to the SM prediction the following one $\mu=1.04\pm 0.20$ \cite{Sopczak:2020vrs,Aaboud:2018zhk,Sirunyan:2018kst}.   Moreover, we have considered the most  recent  and stringent direct constraint for the Higgs-charm Yukawa coupling modifier $\kappa_c$  obtained by CMS and ATLAS  \cite{ATLAS:2022ers,CMS:2022psv}, where  $ 1.1<|\kappa_c|<5. 5$ and this one is interpreted in the $\kappa$-framework \cite{LHCHiggsCrossSectionWorkingGroup:2013rie, LHCHiggsCrossSectionWorkingGroup:2016ypw}. These last constraints  are responsible for the absence of continuity across all allowed regions. This is due to the fact that the Yukawa couplings are strongly sensitive to the $X$ and $Y$ values, hence the BR$(h\to b\bar b)$ is too, as well as BR$(h\to c\bar c)$.  For example, for the Ia scenario, over the region with $2<X<5$,  the BR$(h\to b \bar {b})$ is above the mentioned experimental bounds but, if $Y$ grows larger, $h\to c \bar {c}$ starts to be relevant and  the BR$(h\to b \bar {b})$ decreases until acceptable values. In contrast, in the region $0<X<1$, the channel $h\to b \bar{b}$ is generally inconsistent with experimental data unless $Y$ is small, so that the $h\to c\bar{c}$ decay rate is small too and the BR$(h\to b\bar{b})$ is within the  allowed limits  from the experimental data connected at the CERN machine.

The discussed  event rates are calculated via the formula $\sigma(ep\to \nu_e h j)~\times$ BR$(h\to c\bar c)\times$ 1 ab$^{-1}~\times \epsilon_c^2$ (as mentioned, we take $\epsilon_c=0.24$ as an approximation of the efficiency of a standard $c$-tagging algorithm suitable for the FCC-eh environment \cite{Uta:2014LHeC}). The cross sections and BRs  have been calculated using CalcHEP 3.7.5 \cite{Belyaev:2012qa}, wherein the 2HDM-III has been implemented by ourselves. The proton beam is taken with $50$ TeV of energy ($E_b$), assuming CTEQ6L1 as Parton Distribution Functions (PDFs) \cite{Pumplin:2002vw}, while  the electron beam is considered to be of $60$  GeV ($E_{e^-}$) with a (longitudinal) polarisation ($P_L^{\; e^-}$) of $-80\%$ \cite{Agostini:2020fmq}. %. For each  2HDM-III scenario, we select two specific Benchmark Points (BPs), each corresponding to a maximum and minimum value of $\mu$,  shown in Table \ref{points}. 
For each of these BPs we give herein the common cross section, the BRs into $b\bar b$, $c\bar c$, $s\bar b$ plus Charge Conjugate (C.C.) and $s\bar s$. 

\begin{figure}[ht]
	\centering
		\includegraphics[width=7.8cm]{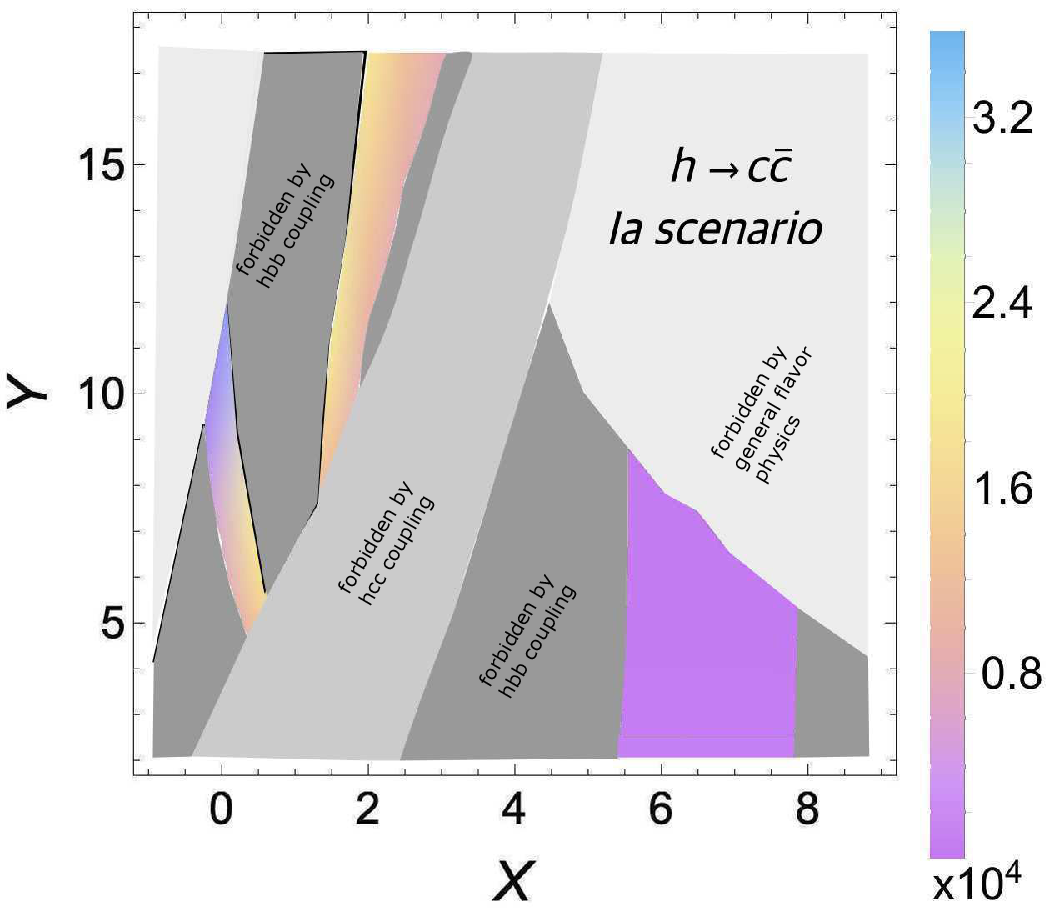}
		\includegraphics[width=7.8cm]{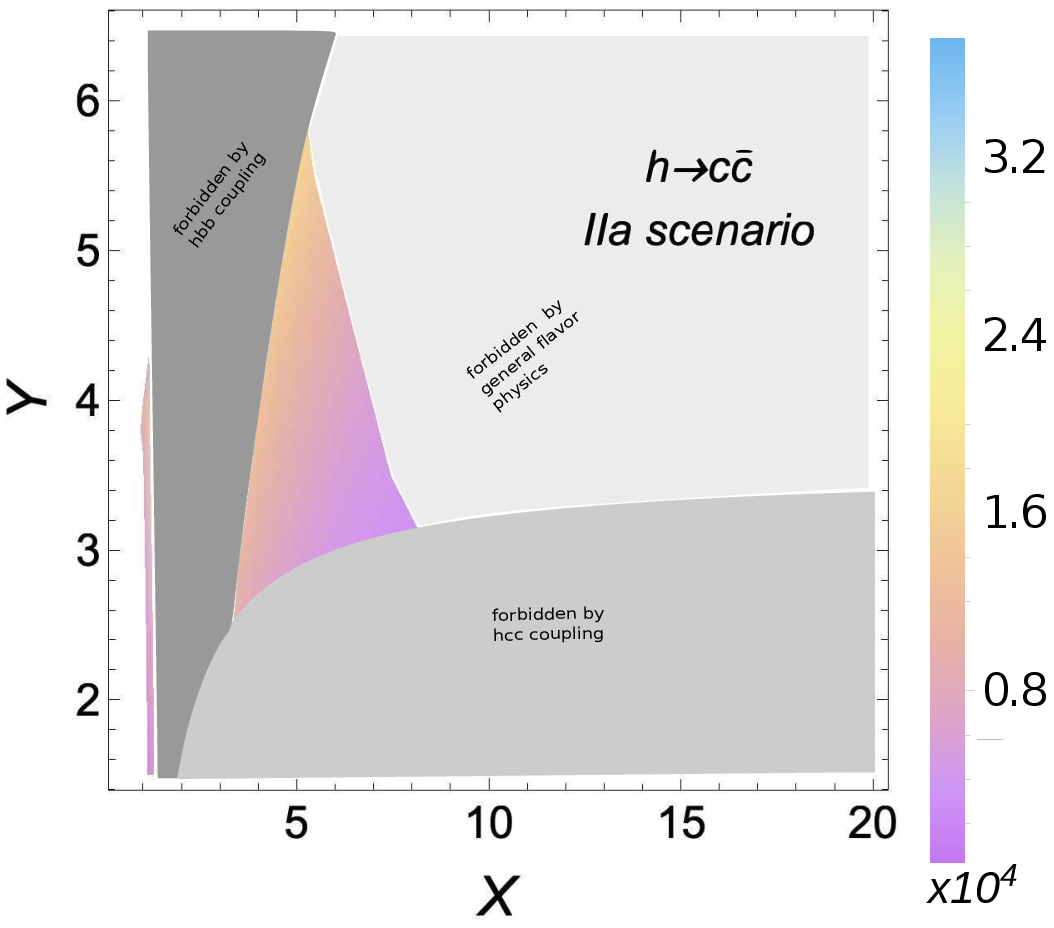}\\
		\includegraphics[width=7.8cm]{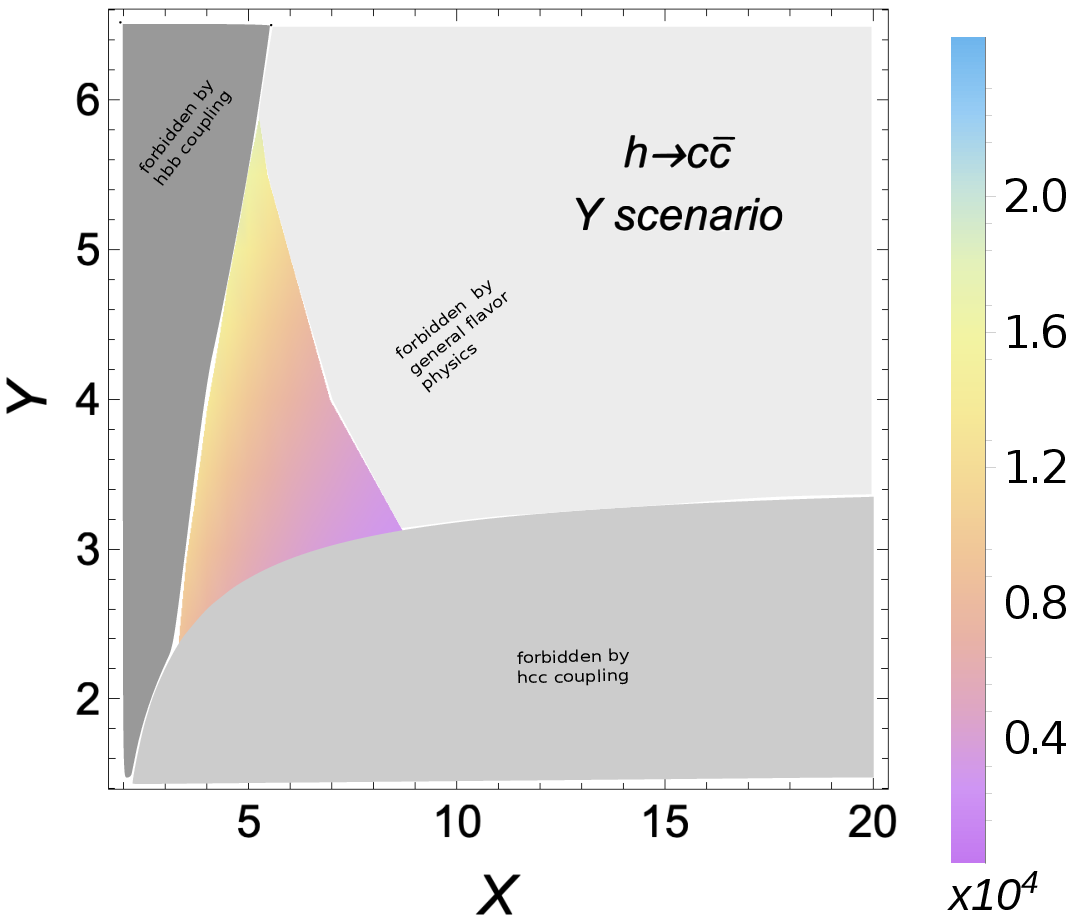}
		\caption{Event rates for each benchmark scenario over the $(X,Y)$ plane computed as  $\sigma(ep\to \nu_e h j)\times {\rm BR}(h\to c\bar{c})\times\epsilon_c^2\times 1 ~{\rm ab}^{-1}$. Here, we have $E_p=50$ TeV and $E_{e^-}=60$ GeV (with $P_L^{\; e^-}=-80\%$).} 
		\label{bench}
\end{figure}

   \begin{table} [h]
	\begin{center}
	\begin{scriptsize}
	\begin{tabular}{|c|c|c|l|c|c|}
		\hline
	 	Point & $X(Z)$ & $Y$ & BR$(\phi^0\to ab)$ & $\sigma(e^-p\to e^- \phi^0 q)$ & Events (1 ab$^{-1}$)\\
		\hline
%		\hline
%		    &              &          & BR$(h\to b\bar{b})=0.696 $ &                      & $2\times 10^5$\\ 
%	    Ia-max & 0.5(0.5) & 4.5    & BR$(h\to c\bar{c})=0.301  $  &  &  $2\times 10^4$						\\ 
%		       &$\mu=1.2$     &                & BR$(h\to s b)=2.7\times 10^{-3} $   &   $0.869$ pb    &$70$ \\ 
%		       &    &                & BR$(h\to s \bar{s})=1.11\times 10^{-8} $   &                    & $0$\\ 
		 %      	        &     &                & BR$(h\to \tau\bar{\tau})=2.9\times 10^{-4} $   &                    & $<10^3$\\ 
		\hline
				    &              &          & BR$(h\to b\bar{b})=0.513 $ &                      & $2\times 10^5$\\ 
	    Ia & 0.5(0.5) & 6.5    &    BR$(h\to c\bar{c})=0.484  $   &      &  $2\times 10^4$\						\\ 
		       &$\mu=0.88$   &           &  BR$(h\to sb)=1.99\times 10^{-3} $ &  $0.875$ pb   & $52$\\ 
	    	       & $\kappa_c=1.5$   &           &  BR$(h\to s\bar{s})=8.18\times 10^{-9} $ &             & $0$\\ 
	%    	       &     &           &  BR$(h\to \tau \tau)=2.96\times 10^{-4} $ &             & $<10^3$\\ 
		\hline
		\hline
	  &              &            & BR$(h\to b\bar{b})=0.67 $ &                          & $2\times 10^5$\\ 
	    IIa & 1(1) & 4 &     BR$(h\to c\bar{c})=0.23  $& $0.958$ pb & $2\times 10^4$ 						\\ 
		       &  $\mu=1.16  $  &     &  BR$(h\to sb)=0.093  $ &                        & $1\times 10^3$\\
		       & $\kappa_c=2$  &                  &  BR$(h\to s\bar{s})=2.87\times 10^{-3}  $ &                        & $7$\\
		%       &     &                  &  BR$(h\to \tau\tau)=7.1\times 10^{-14}  $ &                        & $< 10^3$\\		      
%		\hline
 %&              &            & BR$(h\to b\bar{b})=0.497 $ &                          & $2\times 10^5$\\ 
%	    IIa-min & 5(5) & 5 &     BR$(h\to c\bar{c})=0.288  $& $1.08$ pb & $1\times 10^4$ 						\\ 
%		       &  $\mu=0.86  $  &     &  BR$(h\to sb)=0.208  $ &                        & $3\times 10^3$\\
%		       &     &                  &  BR$(h\to s\bar{s})=1.96\times 10^{-3}  $ &                        & $5$\\
		%       &     &                  &  BR$(h\to \tau\tau)=2.46\times 10^{-3}  $ &                        & $< 10^3$\\		      	
		\hline		
%		\hline
% &              &            & BR$(h\to b\bar{b})=0.702 $ &                          & $2\times 10^5$\\ 
%	    Y-max & 8.5(1/8.5) & 1.5 &     BR$(h\to c\bar{c})=0.031  $& $0.884$ pb & $2\times 10^3$ 						\\ 
%		       &  $\mu=1.21  $  &     &  BR$(h\to sb)=0.264  $ &                        & $7\times 10^3$\\
%		       &     &                  &  BR$(h\to s\bar{s})=2.39\times 10^{-3}  $ &                        & $5$\\
		 %      &     &                  &  BR$(h\to \tau\tau)=6.67\times 10^{-6}  $ &                        & $< 10^3$\\	
		\hline
 &              &            & BR$(h\to b\bar{b})=0.498 $ &                          & $2\times 10^5$\\ 
	    Y-min & 5(-1/5) & 5 &     BR$(h\to c\bar{c})=0.289  $& $1.08$ pb & $2\times 10^4$ 						\\ 
		       &  $\mu=0.86  $  &     &  BR$(h\to sb)=0.21  $ &                        & $7\times 10^3$\\
		       &$\kappa_c=1.7$      &                  &  BR$(h\to s\bar{s})=1.96\times 10^{-3}  $ &                        & $5$\\
		  %     &     &                  &  BR$(h\to \tau\tau)=6.94\times 10^{-4}  $ &                        & $< 10^3$\\	
		\hline		
	\end{tabular}
	\end{scriptsize}
	\caption{Relevant cross sections, BRs and event rates (for the machine configuration given in the previous figure caption)  for our scenarios Ia, IIa and Y, each mapped in terms of $X,Y$ and $Z$ values. %wherein two specific BPs are given in correspondence of two extreme $\mu$ values. 
	We have included the allowed values for 
	 $\mu$ and $\kappa_c$ for each  BPs.  Here, we have included the following tagging efficiencies in the last column: $\epsilon_b=0.6$, $\epsilon_c=0.24$ and $\epsilon_s=0.05$ \cite{Uta:2014LHeC}.}
	\label{points}
	\end{center}
\end{table}	
As  prospect of our work,  for  ee-colliders as ILC \cite{ILC:2019gyn}  ( CLIC \cite{Aicheler:2018arh})  machine the cross sections of Higgs production would be  $\sigma(e^+ e^- \to \nu_e \bar{\nu_e}  h) \sim 220 $ fb  ($\sim 600$ fb) and the main cross section $\sigma(e^- p \to h j \nu_e) \sim 190 $ fb  ($\sim 1000 fb $) for electron proton-colliders LHeC (FCC-he),  with center-of mass energy of  $\sqrt{s}= 1.3 $TeV (  $\sqrt{s}= 3.5 $ TeV) . One can see,  the cross sections are the same order of magnitude. Therefore, the studies of Higgs factories would be complement among ee-colliders and ep- colliders.    

%\begin{table}
%\begin{center}
%\begin{tabular}{|c|c|c|c|c|}
%\hline 
%\sqrt{s}& $\sigma(e^+e^-\to Zh)\times  BR(h\to c\bar c)$&$\sigma(e^+e^-\to 4q)$&$\sigma(e^+e^-\to 2l 2q)$& R=S/B\\ 
%&$ \times BR(Z\to l^+l^-) \ \ [ BR(Z\to qq) ]$& & & \\
%\hline 
%250 GeV & $7.2\times 10^{-3}\ pb$  $[1.2\times 10^{-1}\ pb]$ & $51.2\ pb$ &$1.10 \ pb $ &$6.6\times 10^{-3} [2.3 \times 10^{-3}] $ \\ 
%\hline 
%380 GeV & $3.2\times 10^{-3}\ pb$   $[5.5\times 10^{-2}\ pb]$& $31.1\ pb$ & $1.11\ pb $ &$2.9\times 10^{-3} [1.7 \times 10^{-3}] $\\ 
%\hline 
%500 GeV & $1.7\times 10^{-3}\ pb$   $[2.9\times 10^{-2}\ pb]$&$ 16.2\ pb$ & $1.13\ pb$ &$1.5\times 10^{-3} [1.8 \times 10^{-3}] $\\ 
%\hline 
%1000 GeV & $3.9\times 10^{-4}\ pb$   $[6.6\times 10^{-3}\ pb]$& $4.28\ pb$ & $1.48\ pb$ &$2.6\times 10^{-4} [1.5 \times 10^{-3}] $\\ 
%\hline 
%1500 GeV & $1.7\times 10^{-4}\ pb$  $[2.9\times 10^{-3}\ pb]$ & $2.09\ pb$ & $1.94\ pb$ &$8.8\times 10^{-5} [1.4 \times 10^{-3}] $\\ 
%\hline 
%3000 GeV & $4.1\times 10^{-5}\ pb$  $[7.1\times 10^{-4}\ pb]$& $0.48\ pb$ & $1.92\ pb$ &$2.1\times 10^{-5} [1.5 \times 10^{-3}] $\\ 
%\hline 
%\end{tabular}
%\caption{Raw approximation for signals in a generic $e^+e^-$ collider at center-of-mass energies. Numbers into square parentheses represent a boson $Z$ decaying in two quarks at the final state.  The third and fourth columns show  the irreducible background for four quarks and two quarks two leptons at the final state, respectively.  The R is a simple proportionality reason between signal and background.}
%\label{ee}
%\end{center}
%\end{table} 

\section{Numerical Analysis}
\label{sec:results}
The first step of our numerical analysis is to compare the production and decay rate of signal events to those of the various backgrounds, in presence of acceptance and selection cuts. The latter are implemented at the parton level as 
 $p_T(q)>10$ GeV, $\Delta R(q,q)>0.3$ and $|\eta(q)|<7$, where $q$ represents any quark involved. For our Signal ($S$), we refer to the inclusive rates 
 in Table \ref{points}. For the Background ($B$),   final states of the type $E_T\hspace{-4.5mm} / ~~~+ 3$ jet are considered. In order to  not overload with information the forthcoming histograms, we consider the following five compounded  contributions (wherein $j$ represents any jet except a $b$-one): $\nu 3j$ (it represents the set of $\nu_e j j j$, $\nu_e b j j$ and $\nu_e b b j$ final states),   $\nu_e l l j$ (for any configuration of charged leptons and quarks), $\nu_e t b$, $e3j$ (for $e j j j$, $e b j j$ and $e b b j$) and $e t t$. In  Table \ref{BG}, one can see the corresponding cross sections at parton level  for all these  backgrounds as well as the corresponding event rates for the usual FCC-eh parameters.

\begin{table}[h]
	\begin{center}
		\begin{tabular}{|c|c|c|} 
		\hline
		Background & Cross section [pb] & Number of events \\
		\hline
		\hline
		$\nu_e j j j$ & 172 & $1.75 \times 10^8$   \\
		\hline
		$\nu_e b j j$ & 16.1 & $1.61\times 10^7$  \\
		\hline
		$\nu_e b b  j$ & 1.8 & $1.8\times 10^6$  \\
		\hline
		$\sum \nu 3j$ & 189.9 & $10^8$  \\
		\hline
		\hline
		$\nu_e l l j$ & 3.09& $3.09\times 10^6$  \\
		\hline
%		$\nu_e \tau \tau  j$ & 0.26 & $2.6\times 10^5$ & 25445\\
%		\hline
%		$\sum \nu llj$ & 5.12 & $5.12\times 10^6$ & 418373\\
%		\hline
		\hline
		$\nu_e t b $ & 12.47 & $1.24\times 10^7$  \\
		\hline
		\hline
		$e j j j $ & 948 & $9.48\times 10^8$ \\
		\hline
		$e b j j $ & 17.8 & $1.78	\times 10^7$   \\
		\hline
		$e b b j $ & 75.4 & $75.4 \times 10^7$   \\
		\hline
		$\sum ejjj$ &1040 & $ 10^9$   \\
		\hline
		\hline
		$e t t $ & 0.35 & $3.5\times 10^5$  \\
		\hline
		\end{tabular}
		\caption{Background cross sections and event rates at parton level after the following cuts: $p_T(q)>10$ GeV, $\Delta R(q,q)>0.3$ and $|\eta(q)|<7$ (assuming the usual FCC-eh parameters).}
		\label{BG}
	\end{center}
	\end{table}

For the analysis at detector level, we proceed in the following  way: we use PYTHIA8  \cite{Sjostrand:2014zea} as parton shower and hadronisation generator and Delphes \cite{deFavereau:2013fsa}  as detector simulator. Delphes was run via a FCC-eh card provided in \cite{Uta:2017LHeC}. Finally, we employed MadAnalysis5 \cite{Conte:2012fm} to construct histograms and implement the event selections.

In order to reconstruct the final state of interest, enriched by two $c$-jets, we need to worry about the presence of $b$-jets, as both $c$- and $b$-quarks will originate jets with displaced vertices: thus, just like there is a non-zero probability of $b$-jets being tagged as $c$-jets also the vice versa is possible. Furthermore, a value of $60\%$ for $\epsilon_b$  essentially means that some $b$-jets (precisely, 40\% of them) could be tagged as either $c$-jet or lighter ones\footnote{We neglect here the possibility of $s$-jets to be tagged as $c$- or $b$-ones, so that we need not worry about the role of  $s\bar b$ + C.C. and $s\bar s$ events.}. In essence, it is not obvious what will be the number of true $c\bar c$ events in the  complete di-jet sample (although this is all modelled by Delphes). However, in order to extract the $hc\bar c$ vertex strength,  we can proceed as follows.   To start with, the portion of $b\bar b$ events recognised as such, $N_b\approx \epsilon_b^2$, can be filtered out. Conversely,  around $1-\epsilon_b^2$ of $b\bar b$ events would be accounted as light di-jets ones (including $c\bar c$ ones, that we do not separate out),  which we label as $N_{b\to j}$. (In fact, the latter also includes a $\propto (1-\epsilon_b)$ subleading contribution from mistagged $s\bar b$ + C.C. events.) This will add to the true number of events with only light jets, $N_j$, where $j=s$.  Likewise, the portion of $c\bar c$ events recognised as such is $N_c=\epsilon_c^2$, which in turn implies that $1-\epsilon_c^2$ of these will appear as light jets, labelled as $N_{c\to j}$.    These will also add to the $N_j$ rate alongside the $N_{b\to j}$ one. In order to perform an unbiased measurement of the Yukawa coupling $hc\bar c$ we can only rely on the $N_c$ sample. However, we can use the sample constituted by $N_j+N_{b\to j}+N_{c\to j}$ di-jet events, wherein it is not necessary to extract the fraction of $b$-jets appearing as $c$-jets, for validation purposes, to ensure that the two measurements are consistent with each other, so that, in the remainder of our analysis, we will consider the two cases in parallel. \footnote{This search technique of $h c \bar c$ coupling can be employed for the nearer  future Large Hadron-electron Collider (LHeC).} 

%%%%%%%%%%%%%%%%%%%%%%%%%%%%%%%%

Now, having defined two di-jet samples accounting for flavour (mis)tagging effects, in order to remove the contamination from backgrounds having kinematic configurations similar to the signals,  irrespectively of the flavour composition, we proceed by enforcing the following sets of cuts. (Notice that the kinematics of any $h$ decay into di-jets is the same, given the much larger value of $m_h=125$ GeV with respect to any $m_q$ with $q=d,u,s,c$ and $b$.)  

\begin{itemize}

\item[\textbf{A)}] We impose the following initial conditions for jets and leptons: $p_T(j)>10$ GeV, $p_T(l)>10$ GeV and $|\eta(j)|<6$. Once these requirements are combined with the described tagging procedure,    we have events composed of missing (transverse) energy and three jets.

\item[\textbf{B)}] From the left histogram of  Figure \ref{setb}, one can select the most relevant signature in terms of jet multiplicity,
specifically, we select events with exactly two jets. In fact, the third jet typically  comes directly from the primary vertex and is very forward, when not   outside  the detection zone (i.e., $|\eta(j_3)|>6$), therefore it is not  considered in our analysis. Furthermore, for any jet multiplicity $N[j]>3$, the signal yield is far too depleted to be of numerical interest.  Another cut is over the  missing (transverse) energy, as we take $E_T\hspace {-4.5mm}/~~~ >30$ GeV
(see the histogram on the right-hand side  of Figure \ref{setb}). Specifically, this cut is very strong against the $ejjj$ background, as it keeps only around $20\%$ of such events without penalising the signals excessively.  

\item[\textbf{C)}] The third set of cuts are imposed over the pseudorapidity and transverse momentum of each jet. To start with, we use $\eta$ ordering to tag the first or second jet (i.e., $|\eta(j_1)|>|\eta(j_2)|$). About pseudorapidity, we demand $\eta(j_1)<-3.5$ and $\eta(j_2)<-4$. These cuts are highly  restrictive onto $ejjj$ and $\nu jjj$, keeping around of $8\%$ and $50\%$ of these events, respectively (see top histograms in Figure \ref{eta}).  Furthermore, the selections in jet transverse  momentum are $p_T(j_1)<90$ GeV, which has a strong impact on the $ejjj$ and $ett$ backgrounds, and $p_T(j_2)>30$ GeV, which affects mainly the $\nu jjj$ and $\nu tb$ noises  (see bottom histograms in Figure \ref{eta}). 

\item[\textbf{D)}] We impose that $\Delta R(j_1,j_2)\equiv \sqrt{\Delta\phi(j_1,j_2)^2+\Delta\eta(j_1,j_2)^2}>1.6$. This cut enhances the signal above all backgrounds except $ejjj$:  see Figure \ref{DR1}. 

\item[\textbf{E)}] Finally, we impose a selection on invariant mass for of the two jets, which are the candidates to reconstruct the SM-like Higgs boson mass. Specifically, this cut is $100$ GeV$<M(j_1,j_2)<125$ GeV: see Figure \ref{MAS}.

\end{itemize}
\noindent
(Notice that we have illustrated the kinematics of the Ia incarnation of the 2HDM-III signal but we can confirm that results are extremely similar for the IIa and Y cases as well.)

The response of all signals and backgrounds to each of the above cuts is captured in  Table \ref{resultjj1}. Here, the top value in each row represents the signal rate with no flavour being filtered, i.e., this is the effective di-jet final state defined above as
$N_j+N_{b\to j}+N_{c\to j}$} while the bottom value is the estimated number of $N_c$ events made up by $c\bar c$ pairs recognised as such. It is clear that the kinematic selection is effective in significantly reducing all of the latter without greatly affecting all of the former. This is well exemplified by the  values of the final $S$ versus $B$ rates, including the significances, defined as  $\frac{S}{\sqrt{S+B}}$. The fact that the  corresponding values are always well beyond 5, whichever flavour tagging, clearly indicates the discovery potential of both $h\to jj$ and    $h\to c\bar c$ events at the FCC-eh with a confidence level against the possibility of a background fluctuation far higher than at any hadron collider foreseen at CERN, i.e., a HL-LHC and FCC-hh  \cite{Apollinari:2017lan,Benedikt:2018csr}, and comparable to that of the FCC-ee \cite{Abada:2019zxq}, a future $e^+e^-$ collider therein.

\begin{figure}[ht]
	\centering
		\includegraphics[width=8cm]{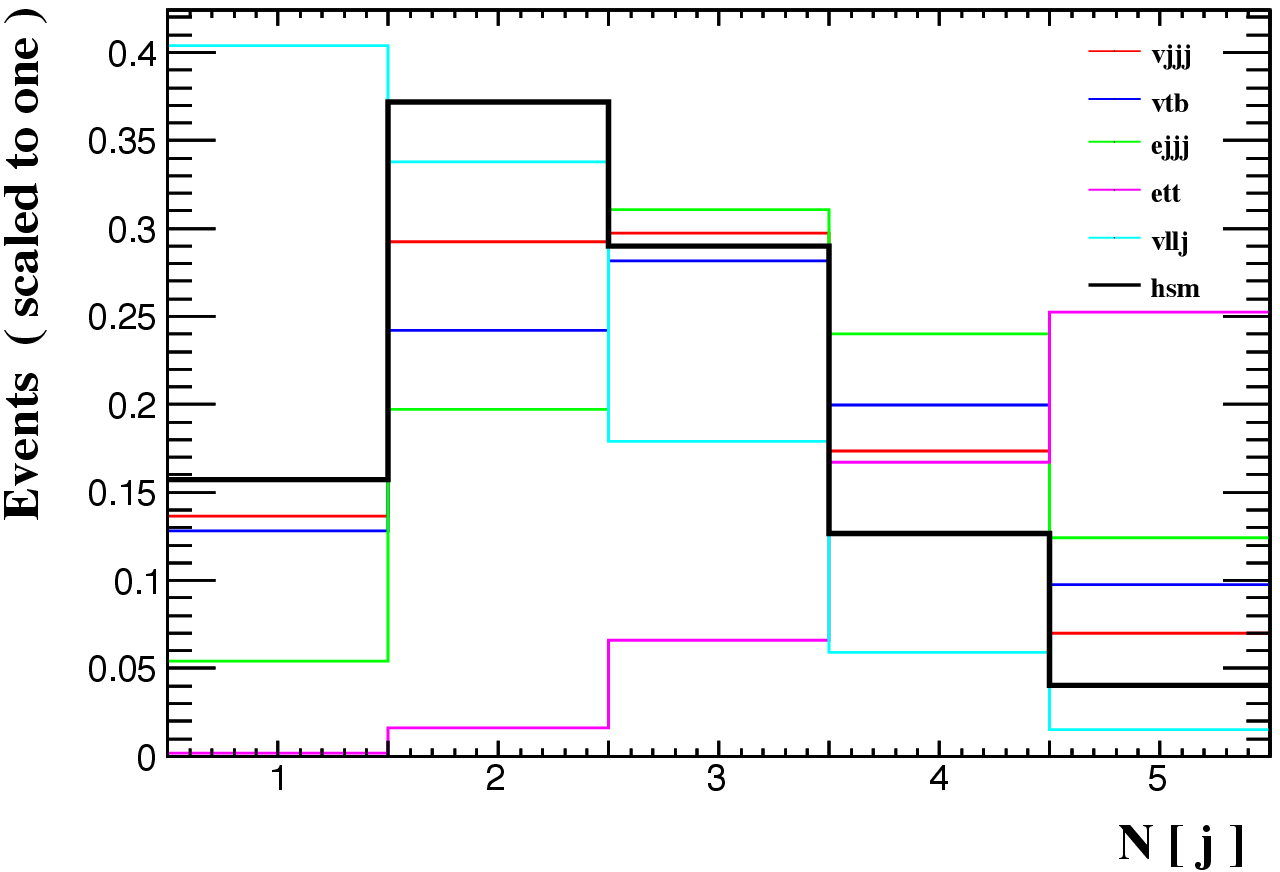}
		\includegraphics[width=8cm]{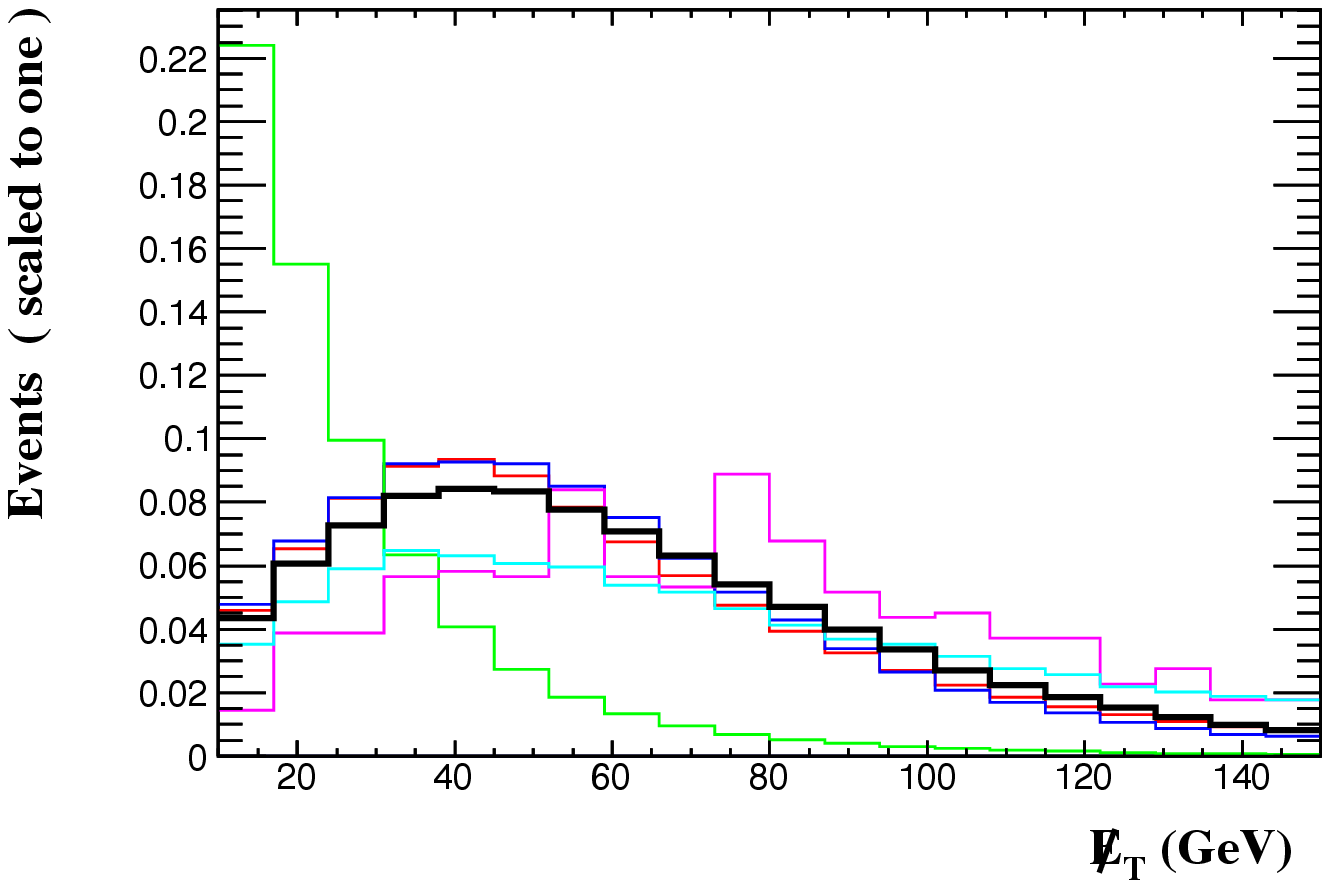}
		\caption{Left) Jet multiplicity (whichever their flavour) distribution. Right) Missing (transverse) energy distribution. These histograms are made for the {Ia} incarnation of the 2HDM-III signal as well as the five categories of background discussed in the text. } 
		\label{setb}
\end{figure}

\begin{figure}[ht]
	\centering
		\includegraphics[width=8cm]{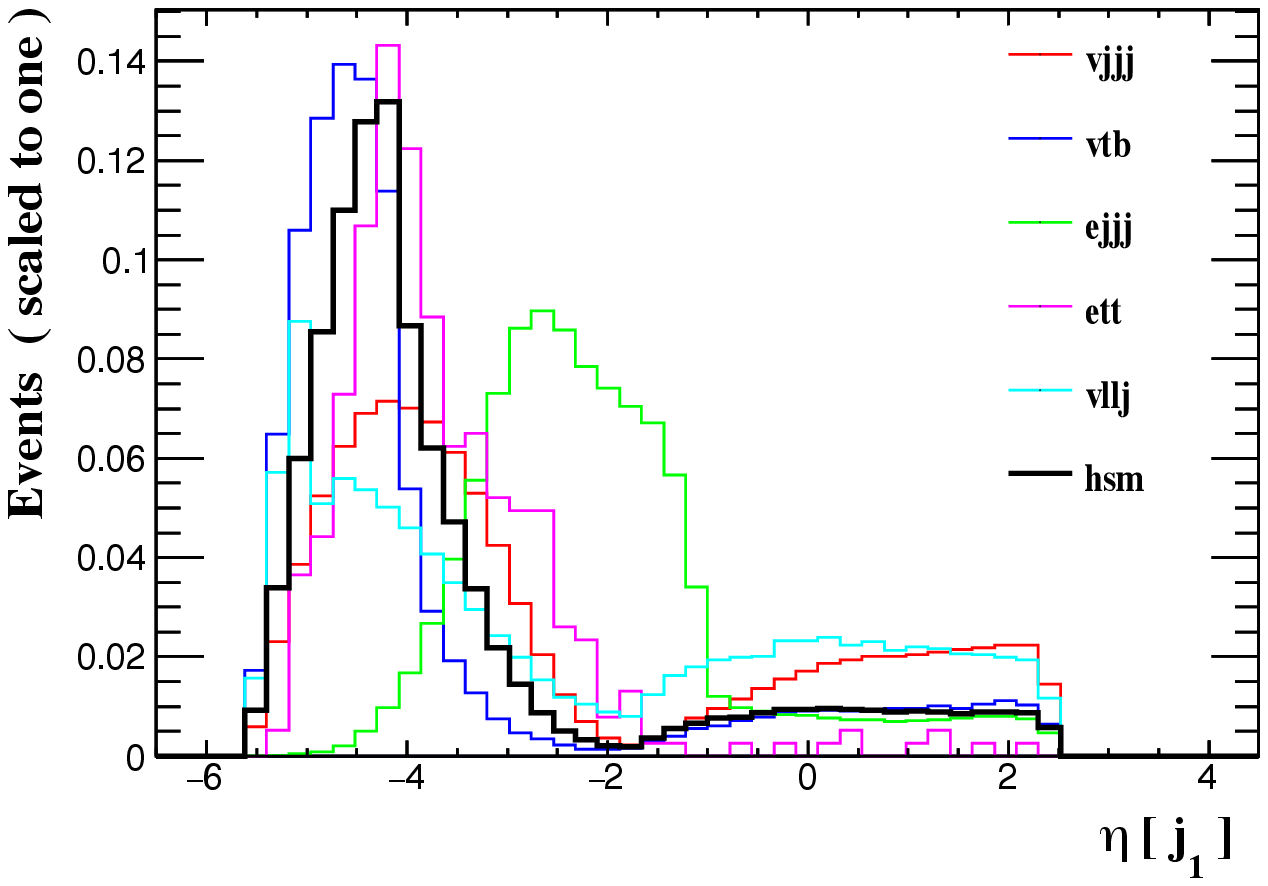}
		\includegraphics[width=8cm]{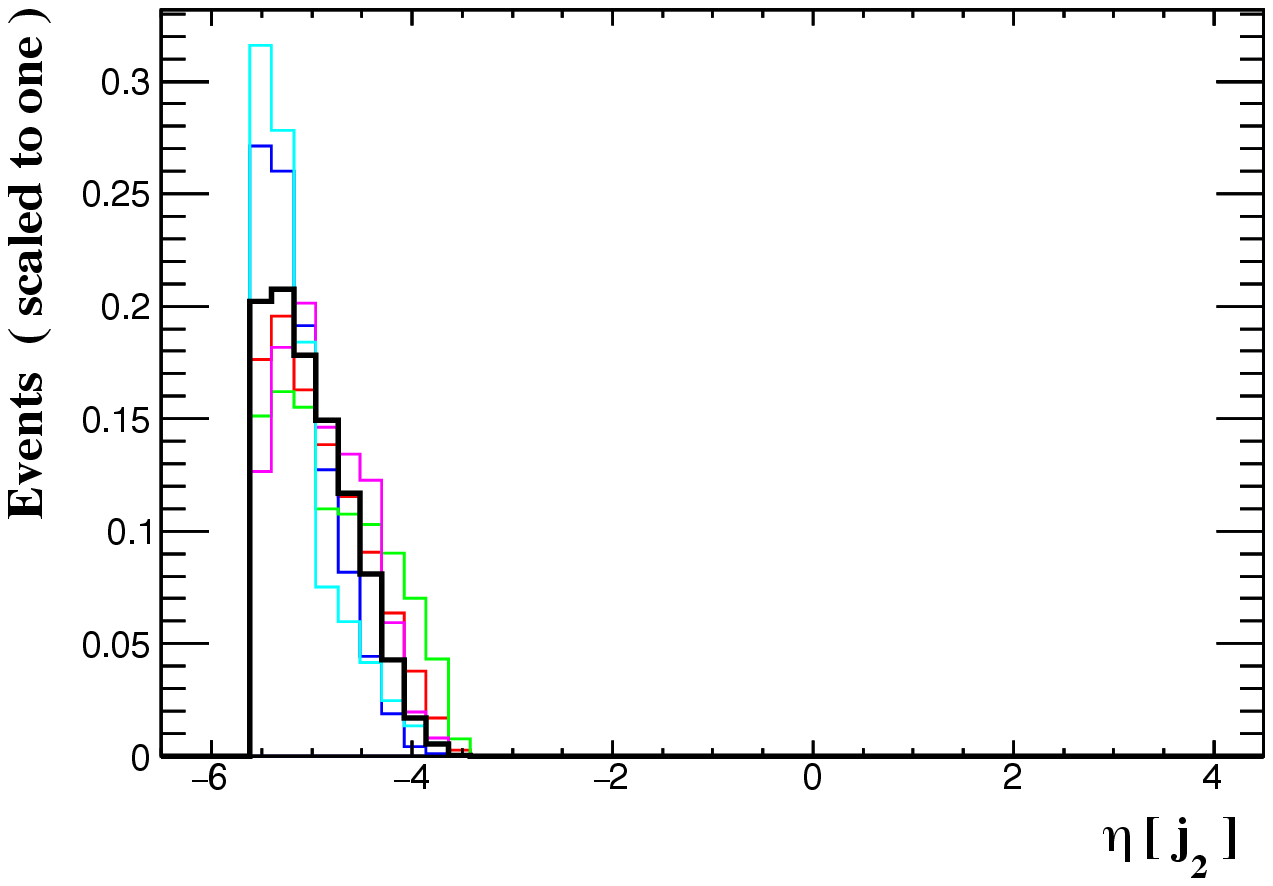}\\
		\includegraphics[width=8cm]{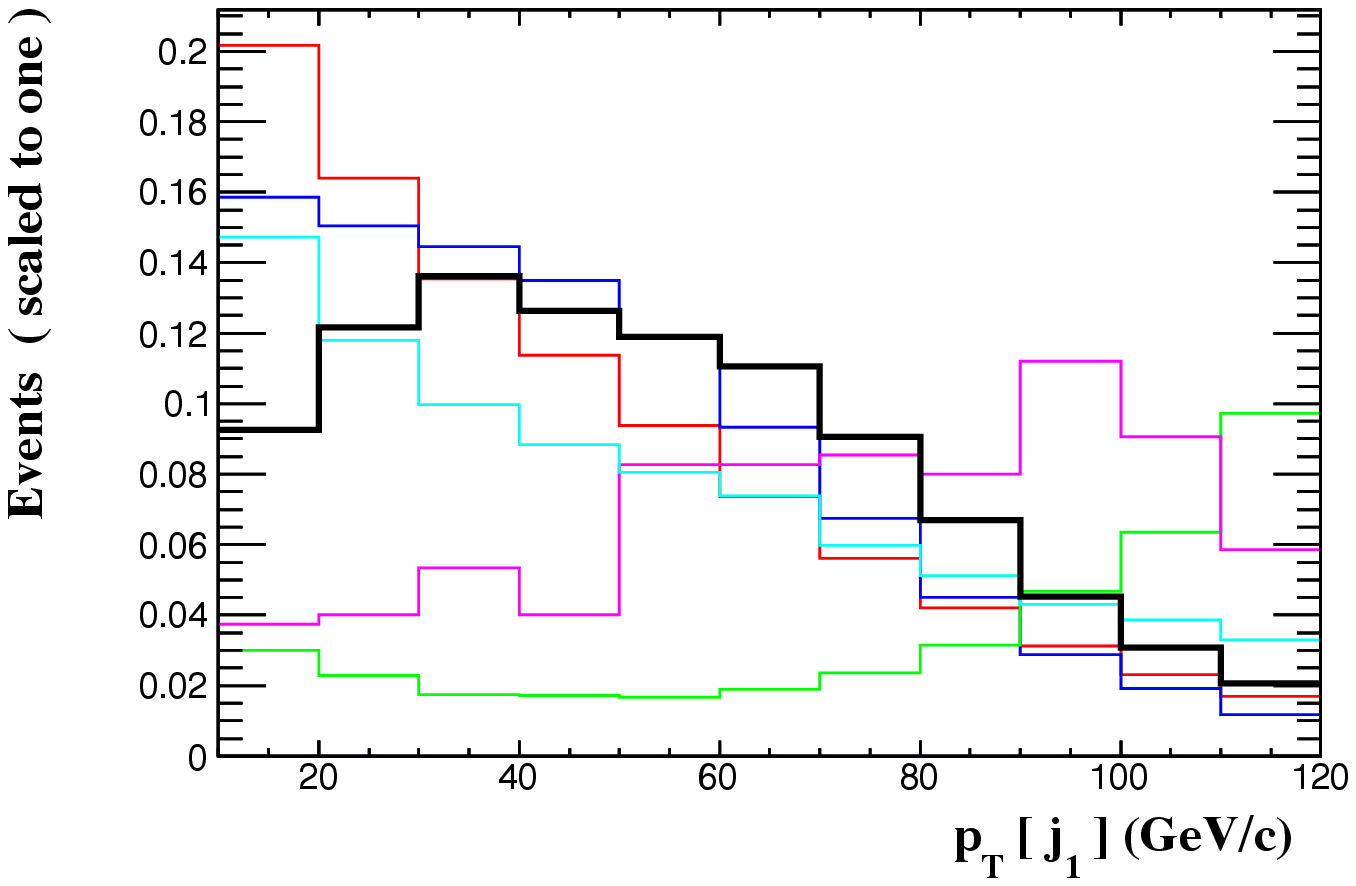}
		\includegraphics[width=8cm]{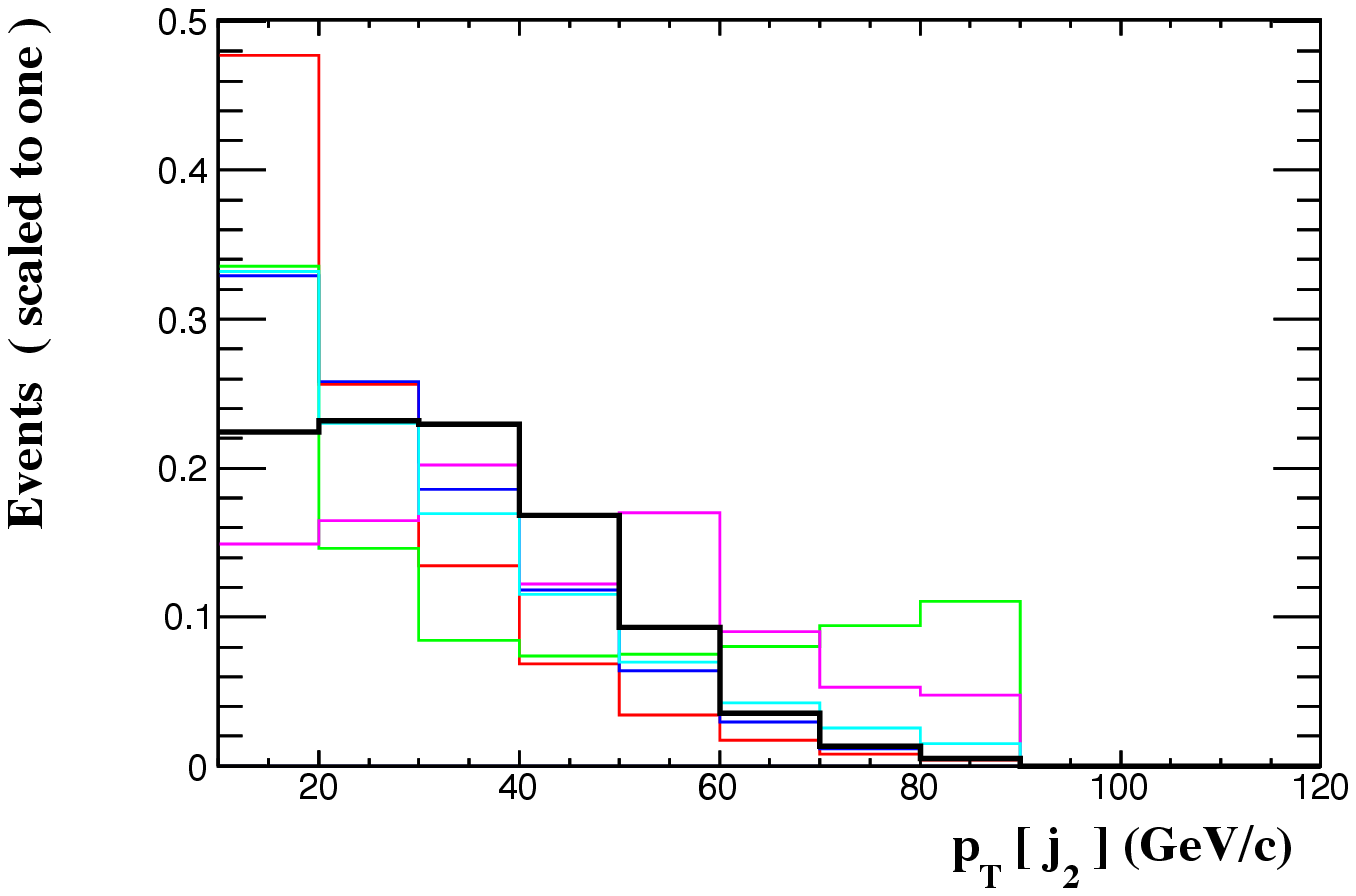}
		\caption{Top-left) Jet pseudorapidity distribution for the first jet. Top-right) Same for the second jet. Bottom-left) Transverse momentum distribution for the first jet. Bottom-right) Same for the second jet. These histograms are made for the {Ia} incarnation of the 2HDM-III signal as well as the five categories of background discussed in the text. }
		\label{eta}
\end{figure}

\begin{figure}[ht]
	\centering
		\includegraphics[width=8cm]{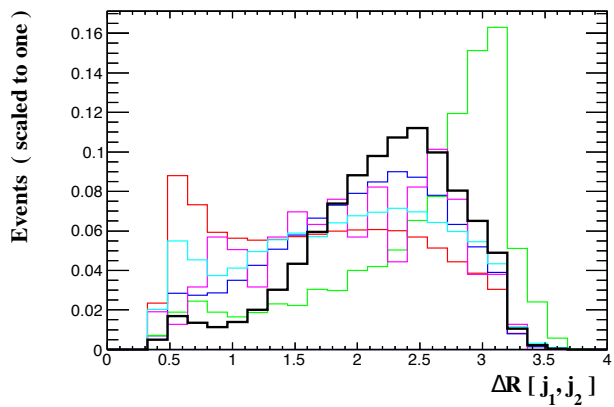}
		\caption{Jet separation distribution. These histograms are made for the {Ia} incarnation of the 2HDM-III signal as well as the five categories of background discussed in the text.} 
		\label{DR1}
\end{figure}

\begin{figure}[ht]
	\centering
		\includegraphics[width=8cm]{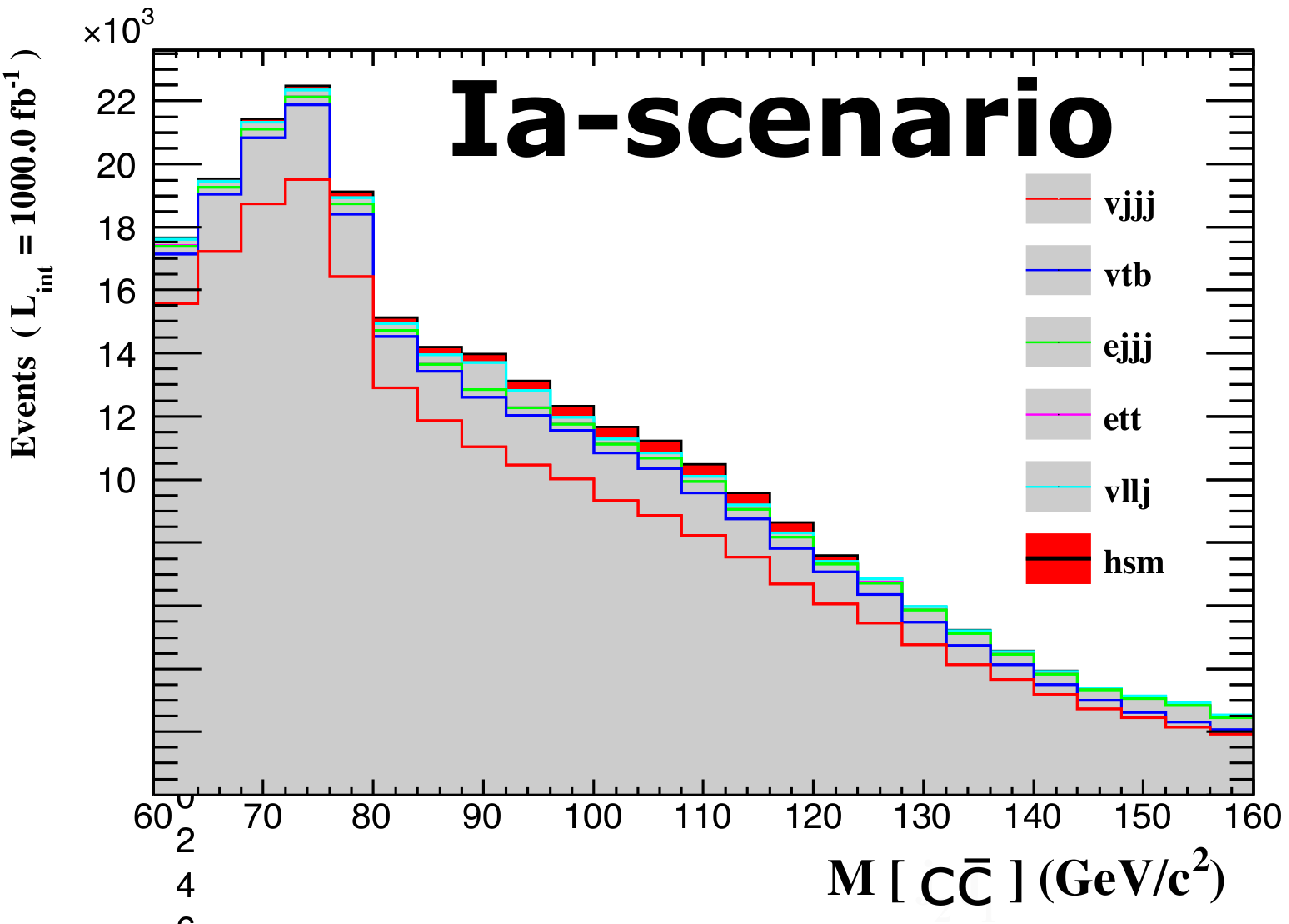}
		\includegraphics[width=8cm]{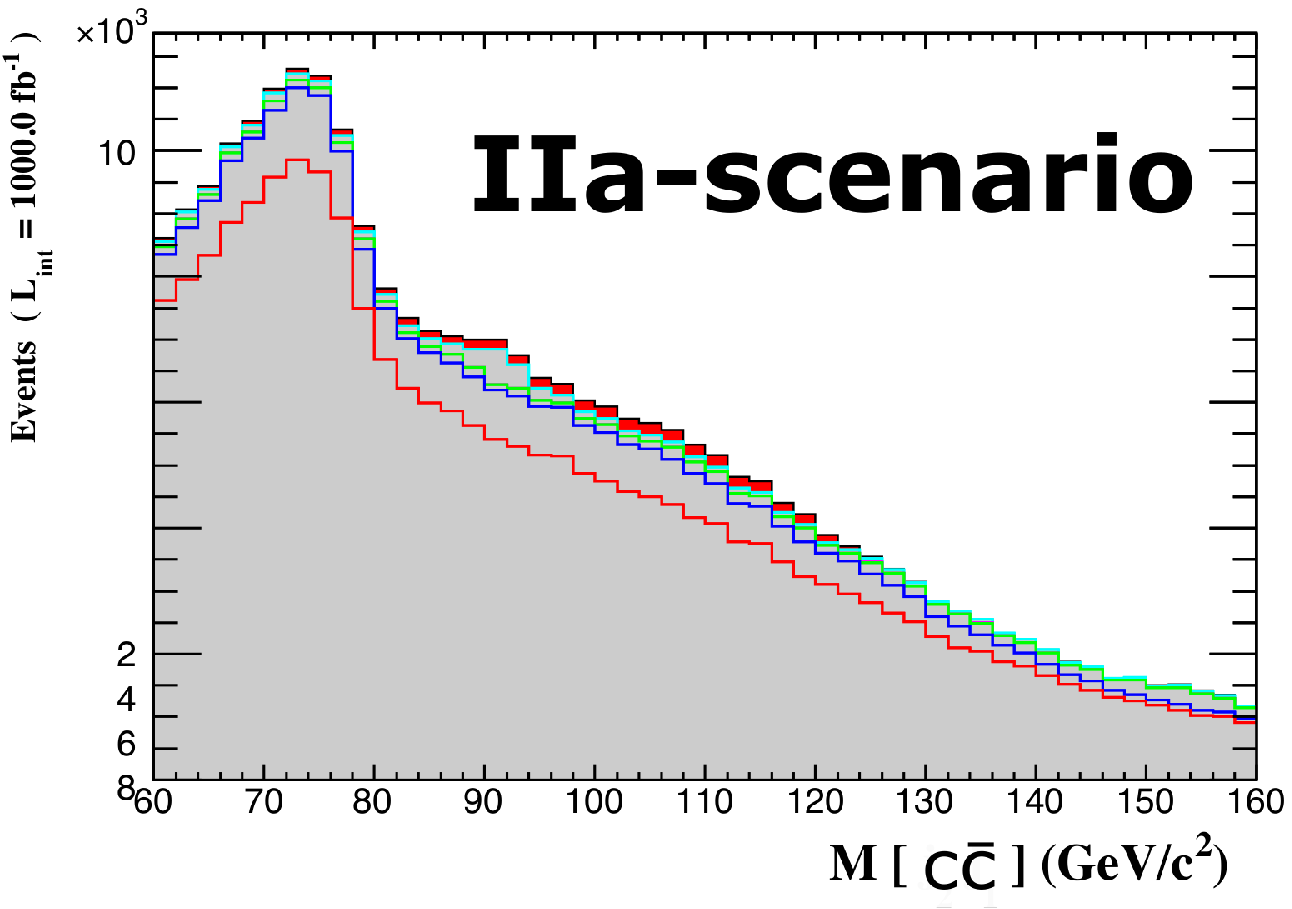}\\
		\includegraphics[width=8cm]{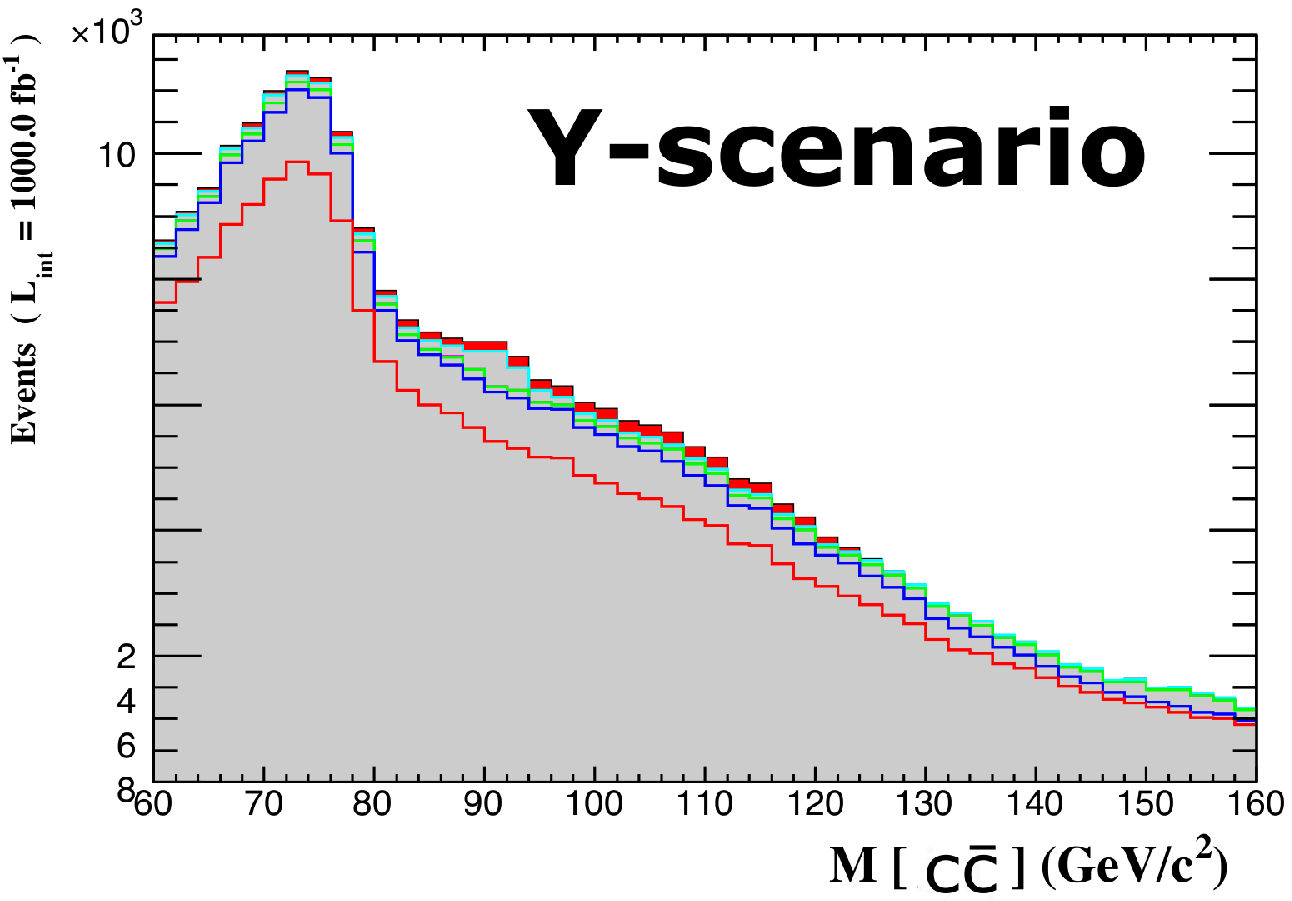}
		\caption{Di-jet invariant mass distribution.  These histograms are made for the {Ia} (top-left), IIa (top-right) and Y (bottom) incarnations of the 2HDM-III signal (red histogram) as well as the five categories of background discussed in the text (here stacked beneath the signal). Here, we present the rates for the case of $c\bar c$-tagged sample.} 
		\label{MAS}
\end{figure}

	 \begin{table} 
	 	\begin{center}
			\begin{tabular}{|l|l|l|l|l|l|l|l||c|}
			\hline
			Signal           &  Raw events  & Sim Events & Set  \textbf{A)}  & Set  \textbf{B)}  &Set  \textbf{C)} &Set \textbf{D)} & Set \textbf{E)}& Significance\\
			\hline
%			\hline
%			Ia-max 	 & 869000 &		885974	&      557172    & 168926  & 82632 & 69576 & 31200 & 31.49\\ 
%						&					&				&		31116	 &  9434  & 4615  &3886	   &1744  &7.08\\
			\hline
			Ia & 875000 &		890530	&      633866    & 190986  & 91117 & 77079 & 36054 & 36.3\\ 
						&					&					&    36075     &  10869  & 5186 & 4387& 2052& 8.31\\
%%%%%%%%%%%%%%%%%%%%%%%%%%%%%%%%%%%%%%%%%%%%%%%%
%%%%%%%%%%%%%%%%%%%%%%%%%%%%%%%%%%%%%%%%%%%%%%%%%%%
			\hline
			IIa 	 & 958000 &		970336	&      609152   & 178088  & 87714 & 72312 & 30898 & 31.19\\ 
						&				&					&    32350	 &  9457  & 4658 & 3840& 1641 & 6.67\\
%			\hline
%			IIa-min & 1080000 &1089956	&    739862    & 209831  & 101811 & 83361 & 36066 & 36.31\\ 
%						&				&					&    42191 	 & 11965  & 5805 & 4754 & 2057 & 8.33\\
%%%%%%%%%%%%%%%%%%%%%%%%%%%%%%%%%%%%%%%%%%%%%%%%%%%
%\hline
%			Y-max 	 & 884000 &		897133	&   508733    & 155513  & 77342 & 64387 & 27283 & 27.59\\ 
%						&				&					&  28056  	& 8576  & 4265 & 3551 & 1505& 6.12\\
			\hline
			Y & 1070000 &1085244	&  736138    & 208665  & 101427 & 83083 & 35824 & 36.08\\ 
						&				&					& 41941 	   & 11884   & 5776  & 4732  & 2040& 8.27\\
%%%%%%%%%%%%%%%%%%%%%%%%%%%%%%%%%%%%%%%%%%%%%%%%%%%
			\hline
			\hline
%%%%%%%%%%%%%%%%%%%%%%%%%%%%%%%%%%%%%%%%%
% V3J
			$\Sigma\nu3j$ & $1.89\times 10^8$  & 19956113  & 176368197 & 40956844   & 9327890 & 4960087 & 820718 & \\  %cb
								&									&					& 10334771&    2399977  & 546593 & 290650  &  48092& \\
%%%%%%%%%%%%%%%%%%%%%%%%%%%%%%%%%%%%%%%%%%%%%%%%%%%%%%
%VTB
			\cline{1-8}
			$\nu t b$      & $1.24\times 10^7$	& 1254485 	 & 7880059 &  1505048  & 759201 & 548492 & 123961 & \\ %cb
								&								&					&  501285  & 95743  & 48296 & 34892 & 7886 &$\Sigma \; B = $ \\
%%%%%%%%%%%%%%%%%%%%%%%%%%%%%%%%%%%%%%%%%%%%%%%%%%%%%%
%e3j
			\cline{1-8}
			$\Sigma e3j$ & $10^9$ 				&104495242   & 73393857 & 3093729   & 29137    & 24770 & 2750 &950207\\ 			
								&								&					&	 52792574  & 2225334  & 20958   & 17817   & 1978&$58865$\\
%%%%%%%%%%%%%%%%%%%%%%%%%%%%%%%%%%%%%%%%%%%%%%%%%%%%%%%%
%ett
			\cline{1-8}
			$e t t$		 & $350000$				&353583  		& 26046 & 380    & 109     & 77 & 21 &\\ %cb
							&									& 					& 14764  & 215    & 62		& 44 & 12&\\
%%%%%%%%%%%%%%%%%%%%%%%%%%%%%%%%%%%%%%%%%%%%%
%vllj
			\cline{1-8} 
			$\Sigma\nu llj$ & $3090000$		& 1434318  & 411923 & 117562   & 29915     & 19052 & 2757 &\\ %cb
									&							&				  &  134029& 38253  & 9733 	   &  6199 & 897&\\
			\hline
			\end{tabular}
		\end{center}
		\caption{Cutflow for all signals and backgrounds. Here,  in each cell, the top line represents the number of light di-jet events while the bottom one refers to those enriched by $c\bar c$ states, as described in the text.}
		\label{resultjj1}
	\end{table}

\section{Conclusions}
\label{sec:conclusions}

In summary, we have studied the process 
$e^- p \to \nu_e h q  $ assuming the decay channel $h \to c \bar{c} $, where $h$ is  the discovered SM-like state,
at a FCC-eh with  $E_b=50$ TeV and $E_{e^-}= 60$ GeV in presence of a $-80\%$ polarisation of the $e^-$ beam. We considered this channel in the context of a 2HDM-III embedding a four-zero texture in the Yukawa matrices and a general Higgs potential, where  both Higgs doublets are coupled with up- and down-type fermions, as a theoretical framework that can be mapped into the standard four types of 2HDM. Hence, we have defined three limits of it reproducing the Type I, II and Y (but not X, which offers no sensitivity to our study) setups. The purpose was to show that this collider has the ability to access the Yukawa coupling between the SM-like Higgs state and $c$-quarks, which can only be determined with significant errors at present and future hadronic machines, like the (HL-)LHC  and FCC-hh. 

Upon accounting for flavour mistagging effects in a realistic way in presence of parton shower, hadronisation and detector effects  and simulating both reducible and irreducible backgrounds, we have proven that large significances can be achieved at such FCC-eh, above and beyond what attainable at the aforementioned hadronic machines and comparable to the FCC-ee expectations. This conclusion applies to all three 2HDM-III incarnations discussed,  each being  exemplified by two BPs at the edges of the currently allowed (by LHC data) interval on the Yukawa coupling between the SM-like Higgs state and $b$-quarks.

\section*{Acknowledgements}
SM is financed in part through the NExT Institute.
 SM also acknowledges support from the UK STFC Consolidated grant ST/L000296/1 and
 the H2020-MSCA-RISE-2014 grant no.  645722 (NonMinimalHiggs). JH-S and CGH have been supported by SNI-CONACYT
(M\'exico), VIEP-BUAP and  PRODEP-SEP (M\'exico)
under the grant `Red Tem\'atica: F\'{\i}sica del Higgs y del
Sabor'.  We all acknowledge useful discussions with Siba Prasad Das.

%%%%%%%%%%%%%%%%%%%%%%%%%%%%%%%%%%%%%%%%%%%%%%%%%%
%	B I B L I O G R A P H Y
%%%%%%%%%%%%%%%%%%%%%%%%%%%%%%%%%%%%%%%%%%%%%%%%%%

\bibliography{cc}
\bibliographystyle{apsrev4-1}

%%%%%%%%%%%%%%%%%%%%%%%%%%%%%%%%%%%%%%%%%%%%%%%%%%
%%%%%%%%%%%%%%%%%%%%%%%%%%%%%%%%%%%%%%%%%%%%%%%%%%
\end{document}